\documentclass[aps,prb,
preprint,
twocolumn,
10pt,
floatfix,
]{revtex4-1}

\usepackage{bm}
\usepackage{amsmath}    
\usepackage{amssymb}    
\usepackage{graphicx}   
\usepackage{booktabs}
\usepackage[hang]{subfigure}

\DeclareMathOperator{\sech}{sech}

\newcommand{\jun}{junction }

\newcommand{\Jos}{Josephson }

\begin{document}

\title[R.Monaco \textit{et al.}]{Confocal Annular Josephson Tunnel Junctions\\ with Large Eccentricity}

\author{Roberto Monaco}
\affiliation{CNR-ISASI, Institute of Applied Sciences and Intelligent Systems ''E. Caianello'', Comprensorio Olivetti, 80078 Pozzuoli, Italy}
\email[Corresponding author's e-mail: ]{r.monaco@isasi.cnr.it}
\author{Jesper Mygind}
\affiliation{DTU Physics, B309, Technical University of Denmark, DK-2800 Lyngby, Denmark}
\email[Author's e-mail: ]{myg@fysik.dtu.dk}
\author{Lyudmila V. Filippenko}
\affiliation{Kotel'nikov Institute of Radio Engineering and Electronics, Russian Academy of Science, Mokhovaya 11, Bldg 7, 125009 Moscow, Russia}
\email[Author's e-mail: ]{lyudmila@hitech.cplire.ru}

\date{\today}
\begin{abstract}
Confocal Annular Josephson Tunnel Junctions (CAJTJs) which are the natural generalization of the circular annular Josephson tunnel junctions, have a rich nonlinear phenomenology due to the intrinsic non-uniformity of their planar tunnel barrier delimited by two closely spaced confocal ellipses. In the presence of a uniform magnetic field in the barrier plane, the periodically changing width of the elliptical annulus generates a asymmetric double-well for a Josephson vortex trapped in a long and narrow CAJTJ. The preparation and readout of the vortex pinned in one of the two potential minima, which are important for the possible realization of a vortex qubit, have been numerically and experimentally investigated for CAJTJs with the moderate aspect ratio $2\!:\!1$. In this work we focus on the impact of the annulus eccentricity on the properties of the vortex potential profile and study the depinning mechanism of a fluxon in more eccentric samples with aspect ratio $4\!:\!1$. We also discuss the effects of the temperature-dependent losses as well as the influence of the current and magnetic noise.
\end{abstract}
\maketitle

\section{Introduction}
\noindent The problem of a particle in a double-well potential (DWP), characterized by two adjacent - in general unequal - potential minima, is almost as old as Quantum Mechanics \cite{hund27}, and one of the first applications was the calculation of an inversion frequency of the ammonia molecule back in 1932 \cite{dennison32}. During the last two decades, the phenomenon of tunneling in asymmetric DWPs was actively considered across several branches of physics and found application in the study of systems, such as the Bose-Einstein condensates in trapped potentials \cite{Levy07,Anker,Albiez,Shin,Hall} and the quantum superconducting circuits based on of low-capacitance \Jos Tunnel Junctions (JTJs) \cite{Makhlin99,Mooij99,Friedman00,You02,Yu02}. The latter have attracted great attention due to their potential use as elementary bits of quantum information (qubits, i.e., two-state quantum-mechanical systems) capable of implementing quantum computing operations.

\begin{figure}[!t]
\centering
\includegraphics[height=6.5cm]{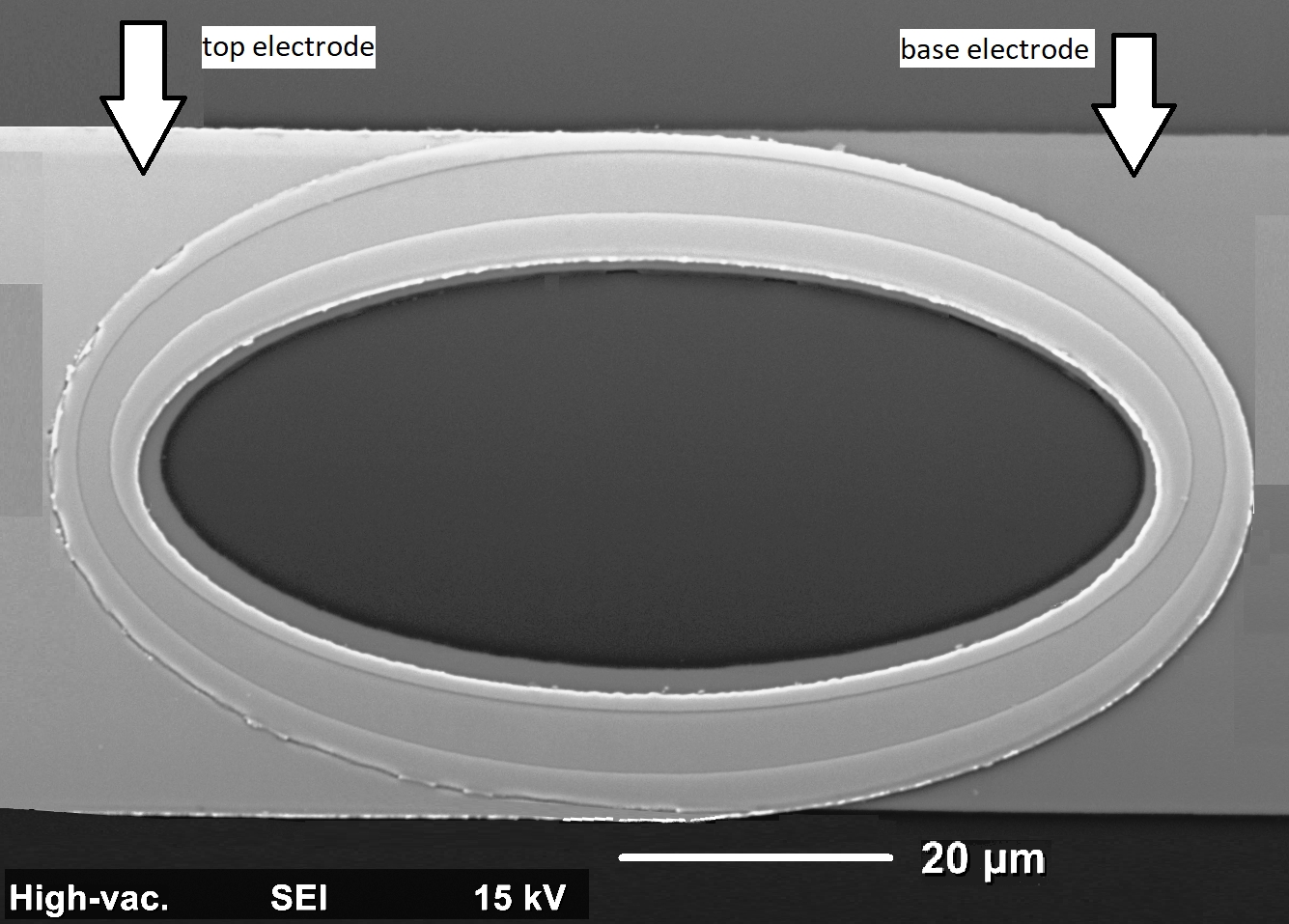}
\caption{Scanning electron microscope image of a confocal annular Josephson tunnel junction (CAJTJ) made of $Nb$ doubly-connected electrodes. The ratio of the major axis and the minor axes is $2\!:\!1$ that implies that the equatorial annulus width is one half of the polar width.}
\label{SEM}
\end{figure}

\noindent Long and narrow, annular JTJs are potential qubit candidates due to their unique capability to trap a Josephson vortex (a supercurrent loop carrying one magnetic flux quantum also called fluxon) whose center of mass becomes the macroscopic collective coordinate of a massive ‘particle’: at sufficiently small temperatures, the particle enters the quantum regime as it shows discrete energy levels within a potential well \cite{wallraff00} and can escape from a potential well via macroscopic quantum tunneling \cite{wallraff03}. Different techniques have been adopted to implement a two-minima fluxon potential in a long JTJ by the application of an external magnetic field and/or on some abrupt changes of the tunnel barrier properties, either the curvature radius \cite{wallraff03} or the Josephson current density \cite{kato96,kim06}. An alternative approach has been proposed \cite{goldobin01} which takes advantage of the proportionality between the fluxon spatial potential and the local width of the long JTJ \cite{nappipagano}. It follows that a large variety of spatially dependent fluxon potentials can be engineered by means of JTLs having a non-uniform width \cite{benabdallah96}. In particular, a magnetically tunable double-well potential was conjectured in a variable-width annular JTL named Confocal Annular Josephson Tunnel Junction (CAJTJ) \cite{JLTP16b} where the tunneling area is delimited by two ellipses having the same focal length. Fig.~\ref{SEM} shows the scanning electron microscope image of a CAJTJ made of $Nb$ doubly-connected electrodes. The CAJTJs represent a generalization of the well-known circular (i.e., zero-eccentricity) annular JTJs intensively studied to experimentally test the perturbation models developed to take into account the dissipative effects in the propagation with no collisions of sine-Gordon kinks \cite{davidson85, dueholm,hue} and  to investigate both the static and the dynamic properties of a fluxon in the spatially periodic potential induced by an in-plane magnetic field \cite{gronbech, ustinov,PRB98}. At variance with the ring-shaped JTLs which have a constant width, it is seen that for the CAJTJs the width of the planar tunnel barrier is smallest at the equatorial point and largest at the poles; the width variation is smoothly distributed along one fourth of the elliptical annulus (mean) perimeter. It is this smooth periodic change of the width of the planar tunnel barrier that makes the physics of CAJTJs very rich and interesting especially since the modeling is very accurate \cite{JLTP16b,JPCM16}. Recently \cite{SUST18}, experiments have shown that a fluxon trapped in a long and narrow CAJTJ experiences a finely tunable DWP and both the preparation and readout of the vortex states in either the left or right state, that are important with respect to the possible realization of a vortex qubit, can be achieved by simple and robust procedures. The previously presented findings concerned CAJTJs with the moderate aspect ratio of $2\!:\!1$ in which, as in Fig.~\ref{SEM}, the (mean) major diameter is twice larger than the minor one. In this paper we investigate the effect of the annulus eccentricity on the properties of the intrinsic fluxon DWP and present both numerical and experimental findings on  CAJTs with aspect ratio of $4\!:\!1$. In particular, we will focus on the mechanisms influencing the depinning of a fluxon from each of the potential wells.

\vskip 5pt
\noindent The paper is organized into five sections. Section II introduces the theoretical framework for the study of a current-biased CAJTJ subjected to an external magnetic field in a modified and perturbed sine-Gordon equation; we then consider the two-minima periodic potential experienced by a trapped fluxon and discuss how the potential changes with the system aspect ratio and how it can be tuned by means of an external in-plane magnetic field and/or bias current. In Sec. III we present numerical simulations of the depinning of a fluxon from each of the two stable states of the DWP in underdamped CAJTJs and describe a protocol to reliably prepare and determine the vortex state. In Sec.IV we  present the experimental data obtained with high-quality low-loss $Nb/Al$-$AlOx/Nb$ window-type CAJTJs in the  presence of in-plane magnetic fields and discuss the role of the temperature and noise on the fluxon depinning. The conclusions of our work are presented in Section V.

\section{Theory of one-dimensional CAJTJs} 

\begin{figure}[t]
\centering
\subfigure[ ]{\includegraphics[width=8cm]{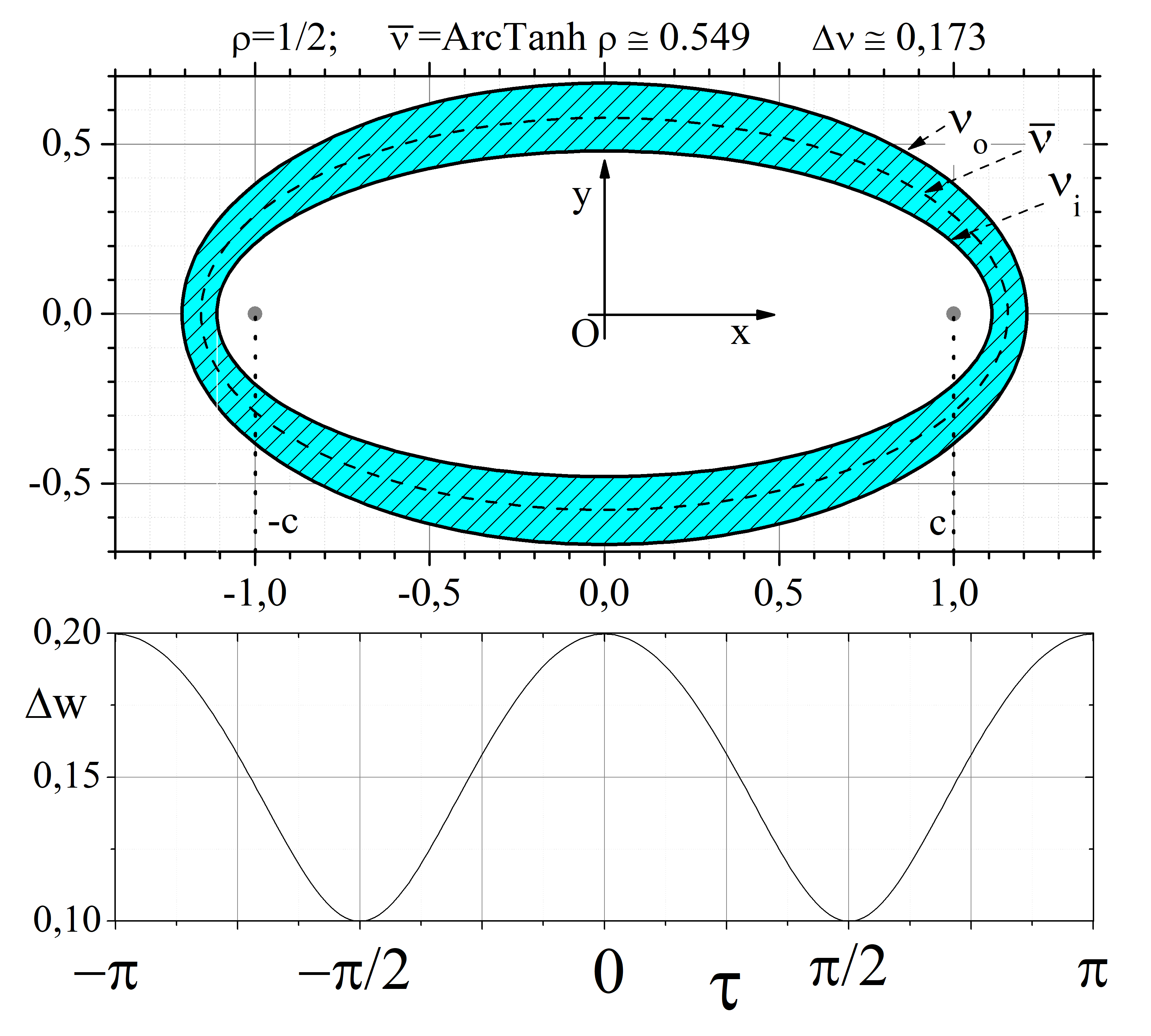}}
\subfigure[ ]{\includegraphics[width=8cm]{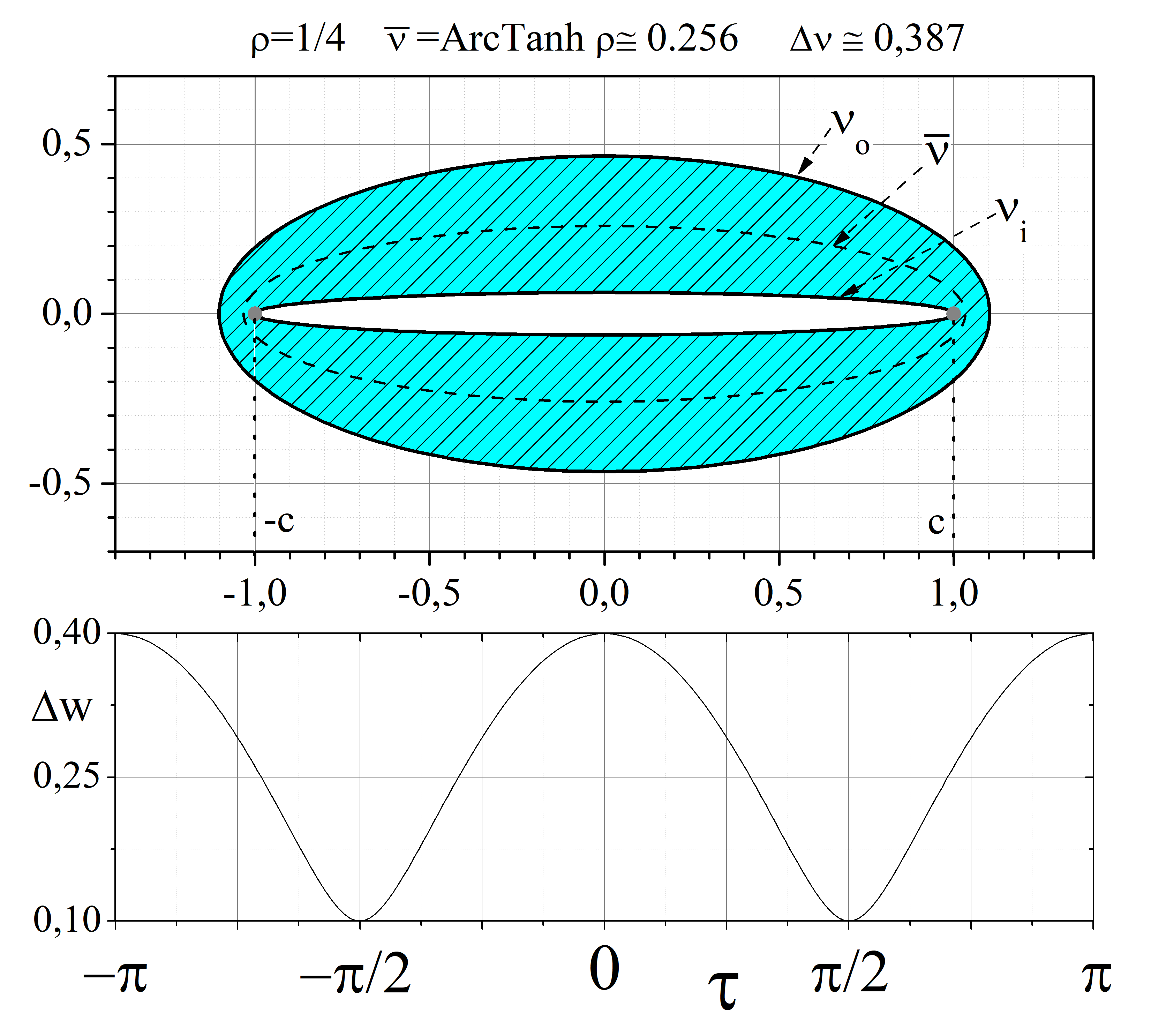}}
\caption{(Color online) Top panels: Drawings of two confocal annuli delimited by closely spaced confocal ellipses (hatched area) identified by the radial elliptic coordinates $\nu_i$ and $\nu_o$; they represent the tunneling areas of two CAJTJs having the same foci - the gray dots at $(\pm 1,0)$ -, but different aspect ratios, $\rho$,: (a) $\rho=1/2$, $\nu_o \approx 0.6358$  and $\nu_i \approx 0.4628$ and (b) $\rho=1/4$, $\nu_o \approx 0.4495$  and $\nu_i \approx 0.0625$. The dashed line locates one more confocal ellipse, called mean or master ellipse as $\bar{\nu}\equiv (\nu_o+\nu_i) /2$. The annuli are built to have the same equatorial or minimum width, $\Delta w_{min}=c \sinh\bar{\nu} \,\Delta\nu=0.1$, with $\Delta\nu\equiv \nu_o-\nu_i$. Bottom panels: The annuli width, $\Delta w$, varies with the angular elliptic coordinate, $\tau$, as given by (\ref{width}).}
\label{ConfAnn}
\end{figure}

\noindent The tunneling area of two CAJTJs with different aspect ratios are sketched by the hatched area in the top panels of Figs.~\ref{ConfAnn}(a) and (b) where the principal diameters of the confocal ellipses are made parallel to the $X$ and $Y$ axes of a Cartesian coordinate system. The common foci (small gray closed circles) lie on the $X$-axis at $(\pm c,0)$. As the focal points move towards the origin, the eccentricity vanishes and the confocal annulus progressively reduces to a circular annulus (with uniform width).


\noindent The geometry of our system suggests the use of the (planar) elliptic coordinate system $(\nu,\tau)$, a two-dimensional orthogonal coordinate system in which the coordinate lines are confocal ellipses and hyperbolae. In this system, any point $(x,y)$ in the $X$-$Y$ plane is uniquely expressed as $(c\cosh\nu\sin\tau, c\sinh\nu\cos\tau)$ with $\nu\geq0$ and $\tau\in[-\pi,\pi]$ for a given positive $c$ value. According to these notations, the origin of $\tau$ lies on the positive $Y$-axis and increases for a clockwise rotation. In the limit $c\to0$, the elliptic coordinates $(\nu,\tau)$ reduce to polar coordinates $(r,\theta)$, where $\theta$ is the angle relative to the $Y$-axis; the correspondence is given by $\tau\to \theta$ and $c\cosh\nu\to r$ (note that $\nu$ itself becomes infinite as $c\to0$). Each possible ellipse with focal points in ($\pm c,0$) is uniquely identified by a value of $\nu$; we will name $\nu_i$ and $\nu_o>\nu_i$ the characteristic values of, respectively, the inner and outer elliptic boundaries of a CAJTJ. Their mean value, $\bar{\nu}= (\nu_o+\nu_i) /2$, labels one more confocal ellipse in between, called mean or master ellipse - see the dashed ellipses in the top panels of Figs.~\ref{ConfAnn}(a) and (b). As the minor and major axes of the master ellipse are given by, respectively, $2c\sinh\bar{\nu}$ and $2c\cosh\bar{\nu}$, we define the aspect ratio of a CAJTJ as $\rho\equiv \tanh\bar{\nu}$ and its (mean) eccentricity as $e^2 \equiv 1-\rho^2=\sech^2\bar{\nu}$. 


\vskip 5pt
\noindent For closely spaced inner and outer ellipses, $\Delta\nu\equiv \nu_o-\nu_i<<1$, the expression of the local annulus width is \cite{JLTP16b}:
\begin{equation}
\Delta w(\tau)=c\mathcal{Q}(\tau)\,\Delta\nu,
\label{width}
\end{equation}

\noindent where $\mathcal{Q}(\tau)$ is the elliptic scale factor defined by $\mathcal{Q}^2(\tau) \equiv \sinh^2\bar{\nu} \sin^2\tau+\cosh^2 \bar{\nu} \cos^2 \tau= \sinh^2\bar{\nu}+ \cos^2\tau=\cosh^2\bar{\nu} - \sin^2\tau=(\cosh2\bar{\nu} + \cos2\tau)/2$. The width of the confocal annulus is smallest at the equatorial points, with $\Delta w_{min}=c \Delta \nu  \sinh \bar{\nu}$, and largest at the poles, with $\Delta w_{max}=c \Delta \nu  \cosh \bar{\nu}$; then $\Delta w_{min}=\rho \Delta w_{max}$, i.e., interestingly, as we make the confocal annulus more eccentric we enhance the width spread $\Delta w_{max}/\Delta w_{min}$. This is clearly seen in  Fig.~\ref{ConfAnn} where the annuli are made to have the same equatorial widths, $\Delta w_{min}=0.1$: as we halve the aspect ratio, $\rho$, passing from $1/2$ to $1/4$, the polar width, $\Delta w_{max}$, doubles from $0.2$ to $0.4$. The annuli width variations with the angular elliptic coordinate, $\tau$, are shown in the bottom panels of Figs.~\ref{ConfAnn}(a) and (b) as given by (\ref{width}).

\vskip 5pt
\noindent In the small width approximation, $\Delta w_{max}<< \lambda_J$, where $\lambda_J$, called the \Jos penetration length, gives a measure of the distance over which significant spatial variations of the \Jos phase occur, the system becomes one-dimensional. It has been derived that the $\nu$-independent \Jos phase, $\phi(\tau,\hat{t})$, of a narrow CAJTJ in the presence of a spatially homogeneous in-plane magnetic field ${\bf H}$ of arbitrary orientation, $\bar{\theta}$, relative to the $Y$-axis, obeys a modified and perturbed sine-Gordon equation with a space dependent effective Josephson penetration, $\lambda_J/Q(\tau)$,  length inversely proportional to the local junction width \cite{JLTP16b}:
\begin{equation}
 \left[\frac{\lambda_J}{c\,\mathcal{Q}(\tau)}\right]^2 \left(1+\beta\frac{\partial}{\partial \hat{t}}\right) \phi_{\tau\tau} - \phi_{\hat{t}\hat{t}}-\sin \phi =\alpha \phi_{\hat{t}} - \gamma(\tau) + F_h(\tau),
\label{psge}
\end{equation}

\noindent where $\hat{t}$ is the time normalized to the inverse of the so-called (maximum) plasma frequency, $\omega_p$. The critical current density, $J_c$, was assumed to be uniform. The subscripts on $\phi$ are a shorthand for derivative with respect to the corresponding variable. Furthermore, $\gamma(\tau)=J_Z(\tau)/J_c$ is the local normalized density of the bias current and 
\begin{equation}
F_h(\tau)\equiv h\Delta \frac{\cos\bar{\theta}\cosh\bar{\nu} \sin\tau-\sin\bar{\theta}\sinh\bar{\nu}\cos\tau }{\mathcal{Q}^2(\tau)}
\label{Fh}
\end{equation}
\noindent is an additional forcing term proportional to the applied magnetic field; $h\equiv H/J_c c$ is the normalized field strength for treating long CAJTJs and $\Delta$ is a geometrical factor which sometimes has been referred to as the coupling between the external field and the flux density of the annular junction \cite{gronbech}. As usual, the $\alpha$ and $\beta$ terms in (\ref{psge}) account for, respectively, the quasi-particle shunt loss and the surface losses in the superconducting electrodes. The perimeter of the master ellipse is $L= 4c \cosh\bar{\nu}\,\texttt{E}(e^2)$, where $\texttt{E}(e^2)\equiv \texttt{E}(\pi/2,e^2)$ is the {\it complete} elliptic integrals of the second kind of argument $e^2$. Then the normalized or electric length, $\ell=L/\lambda_J$, of the CAJTJ of a given aspect ratio grows linearly with the foci distance, $2c$. 

\vskip 5pt
\noindent When cooling an annular JTL below its critical temperature one or more flux quanta may be trapped in its doubly connected electrodes. The algebraic sum of the flux quanta trapped in each electrode is an integer number $n$, called the winding number, counting the number of Josephson vortices (fluxons) trapped in the \jun barrier. To take into account the number of trapped fluxons, (\ref{psge}) is supplemented by periodic boundary conditions \cite{PRB96,PRB97}:
\vskip -8pt
\begin{subequations}
\begin{eqnarray} \label{peri1}
\phi(\tau+2\pi,\hat{t})=\phi(\tau,\hat{t})+ 2\pi n,\\
\phi_\tau(\tau+2\pi,\hat{t})=\phi_\tau(\tau,\hat{t}).
\label{peri2}
\end{eqnarray}
\end{subequations}
\vskip -4pt

\subsection{The single fluxon potential}

\noindent In the absence of dissipative and driving forces, the simplest topologically stable dynamic solution of (\ref{psge}) on an infinite line, in a first approximation, is a $2\pi$-kink (single fluxon) centered at a time-dependent coordinate $s_0(\hat{t})$ and moving with instantaneous (tangential) velocity $\hat{u}\equiv d({s}_0/{\lambda_J})/d\hat{t}=(c/\lambda_J)\mathcal{Q}(\tau_0) d\tau_0/d{\hat{t}}$:

\begin{equation}
\tilde{\phi}(\tau,\hat{t})= 4 \arctan \exp \left\{\wp[s(\tau)-s_0(\hat{t})]/\lambda_J \right\},
\label{tilde}
\end{equation}

\noindent where $\wp=\pm1$ is the topological charge, i.e., the fluxon polarity \cite{scott} and $s(\tau)$ the non-linear curvilinear coordinate $s(\tau)=c \int_{0}^{\tau} \mathcal{Q}(\tau')d\tau'$. Inserting the phase profile in (\ref{tilde}) into (\ref{psge}) it was derived that \cite{JPCM16}, in the absence of external forces, the energy of a non-relativistic fluxon ($\hat{u}<<1$), $\hat{E}= \hat{K}+\hat{U}_w$, is conserved. The circumflex accents denotes normalized quantities. $\hat{E}$ is normalized to the characteristic energy, $\mathcal{E}=\Phi_0 J_c \lambda_J c \Delta \nu/2\pi$. Both the kinetic energy, $\hat{K}(\tau_0)\approx 4 \mathcal{Q}(\tau_0) \hat{u}^2$, and the intrinsic potential energy, $\hat{U}_w(\tau_0) \approx 8 \mathcal{Q}(\tau_0)$, are position dependent through the scale factor $\mathcal{Q}$, that is, in force of (\ref{width}), they are proportional to the annulus width. This is consistent with the relativistic expression $\hat{E}=\hat{m}(\tau_0)/\sqrt {1-\hat{u}^2(\tau_0)}$ reported by Nappi and Pagano \cite{nappipagano}, provided that we introduce the position dependent reduced rest mass $\hat{m}(\tau_0)=8 \mathcal{Q}(\tau_0)$ of the fluxon. Note that the energy, $E_0$, of a CAJTJ containing one static vortex is $8\mathcal{E}\sinh\bar{\nu}$. With experimentally accessible geometrical and electrical parameters the normalizing energy is much larger than $k_B T$ - $\mathcal{E}=O(10^4 K)$ - and the fluxon rest mass, $m_0 \equiv E_0/\bar{c}^2$, happens to be much smaller than the electron rest mass $m_e$ - $m_0=O(10^{-3} m_e)$ -, where the so-called Swihart velocity \cite{Swihart}, $\bar{c}$, is the characteristic velocity of the electromagnetic waves in JTJs.

\medskip
As can be discerned from the plots in the bottom panels of Figs.~\ref{ConfAnn}(a) and (b), $\hat{U}_w \propto \Delta w$ expresses a $\pi$-periodic potential energy function independent on the fluxon polarity, and uniquely determined by the CAJTJ aspect ratio. The potential wells are located at $\tau_0=\pm \pi/2$, where the confocal annulus is narrowest. The left $|L\rangle$ and right $|R\rangle$ wells of the potential constitute stable classical states for the vortex with degenerate ground state energy. Considering that $\sinh\bar{\nu}\leq \mathcal{Q}(\tau) \leq \cosh\bar{\nu}$, the potential wells are separated by a normalized energy barrier, $\Delta\hat{U}_{w}\equiv \hat{U}_{w,max}-\hat{U}_{w,min}=8\exp-\bar{\nu}$, uniquely determined by the system aspect ratio. As an example, the change of the aspect ratio from $2\!:\!1$ to $4\!:\!1$ in Figs.~\ref{ConfAnn} results in the triplication of the energy barrier (and so of the potential gradient).

\begin{figure}[t]
\centering
\includegraphics[width=8cm]{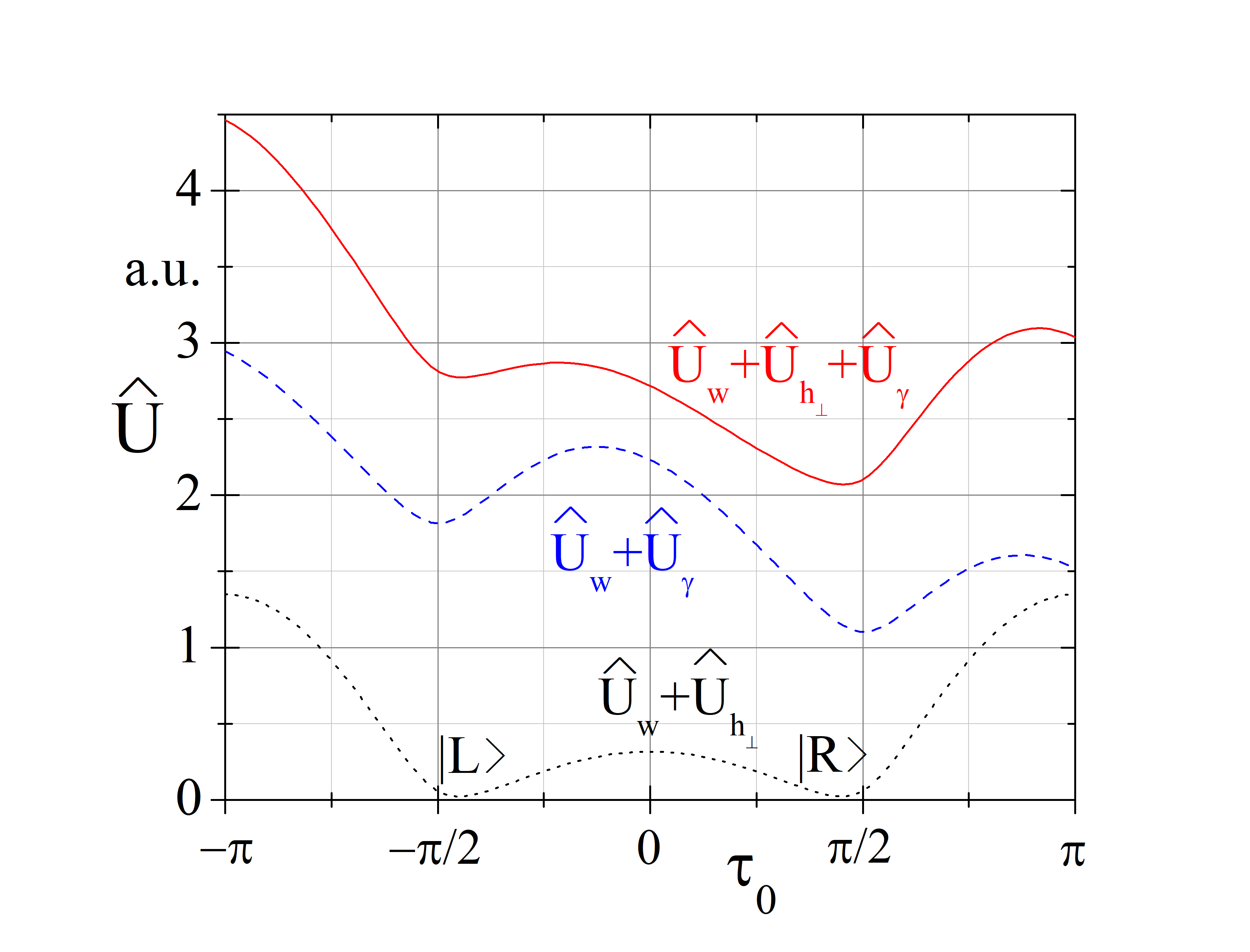}
\caption{(Color online) Schematic representation of the fluxon potential under different conditions. The dotted line at the bottom refers to the symmetric double-well potential $\hat{U}_w+\hat{U}_{h_\bot}$ in the presence of a uniform in-plane magnetic field perpendicular to the long annulus diameter with two minima at $\tau_0 \approx \pm \pi/2$ coincident with the degenerate states $|R\rangle$ and $|L\rangle$; the dashed line corresponds to $\hat{U}_w+\hat{U}_\gamma$ and displays the tilting of the potential due to a uniform bias current; the solid line shows the asymmetric potential $\hat{U}_w+\hat{U}_{h_\bot}+\hat{U}_\gamma$ in the more general case of applied (perpendicular) magnetic field and bias current. The three potentials are shifted by arbitrary vertical offsets.}
\label{potentials}
\end{figure}

\medskip
\noindent In the presence of small applied magnetic field and bias current, two more terms contribute to the total potential energy, $\hat{U}$, experienced by the fluxon:
\vskip -8pt
\begin{equation}
\hat{U}(\tau_0)=\hat{U}_w(\tau_0)+\hat{U}_h(\tau_0)+\hat{U}_\gamma(\tau_0).
\label{Utot}
\end{equation}

\noindent $\hat{U}_h(\tau_0)$ is the magnetic potential such that $dU_h /d\tau= 2\pi \wp (\lambda_J/c) \mathcal{Q}^2(\tau) F_h(\tau)$, i.e.,:
$$\hat{U}_h(\tau_0)=\hat{U}_{h_\bot}(\tau_0)+\hat{U}_{h_{||}}(\tau_0)\approx$$
\begin{equation}
\approx 2\pi \wp (\lambda_J/c)  \Delta \left( h_\bot \cosh\bar{\nu} \cos\tau_0 + h_{||} \sinh\bar{\nu}\sin\tau_0 \right),
\label{uh}
\end{equation}

\noindent where $h_\bot\equiv h \cos\bar{\theta}$ and $h_{||}\equiv h \sin\bar{\theta}$ are the components of the in-plane magnetic field, respectively, perpendicular and parallel to the CAJTJ's major diameter. $\hat{U}_h(\tau)$ is $2\pi$-periodic and $\pi$-antiperiodic in $\tau$, i.e., $\hat{U}_h(\tau+\pi) =-\hat{U}_h(\tau)$, then it averages to zero over one period. It is important to note that $\hat{U}_{h_\bot}$ is even in $\tau_0$ as the intrinsic potential $\hat{U}_w$, while $\hat{U}_{h_{||}}$ is odd and breaks the system parity.

\noindent Furthermore, $\hat{U}_\gamma(\tau_0)$ is the current-induced potential such that $d\hat{U}_\gamma /d\tau=2\pi \wp (\lambda_J/c) \mathcal{Q}^2(\tau) \gamma(\tau) $; assuming a uniform current distribution $\gamma(\tau)=\gamma_0$, it is:

\vskip -8pt
\begin{equation}
\hat{U}_\gamma(\tau_0)\equiv \pi \wp (\lambda_J/c) \gamma_0 \left( \tau_0 \cosh2\bar{\nu} + \frac{1}{2}\sin2\tau_0\right).
\label{ugamma}
\end{equation}
\vskip -4pt

\noindent  Fig.~\ref{potentials} qualitatively explains how the width-dependent fluxon potential can be tuned by means of an externally applied magnetic field and/or a bias current. The dotted curve at the bottom of Fig.~\ref{potentials} shows the fluxon potential in the presence of a (negative) perpendicular magnetic field $h_\bot$; the potential $\hat{U}_w+\hat{U}_{h_\bot}$ is still invariant under a parity transformation $(\tau_0 \to -\tau_0)$ and develops into a field-controlled symmetric potential with finite walls and two spatially separated minima. Increasing the amplitude of the magnetic field, eventually the minima coalesce and the perturbed potential becomes single-welled for a (perpendicular) threshold field strength $h_{\bot}^{*}$. The evolution of the potential $\hat{U}$ with an increasing parallel field, $h_{||}$, follows a quite different pattern (not shown in Fig.~\ref{potentials}), but again a threshold value, $h_{||}^{*}$, exists where the potential $\hat{U}_w+\hat{U}_{h_{||}}$ becomes single-welled. It means that the inter-well barrier can be fine-tuned and made arbitrarily small by means of both a perpendicular or a parallel in-plane magnetic field. The dashed line in Fig.~\ref{potentials} shows the total potential when a bias current is feeding the CAJTJ. The resulting potential, $\hat{U}_w+\hat{U}_{\gamma}$, is qualitatively similar to the well-studied tilted washboard potential for the phase difference of a small JTJ biased below its critical current; the only difference is that in our case the degree of freedom is the spatial coordinate, rather than the Josephson phase difference. Indeed, the potential profile can be tilted either to left or to right depending on the polarity of the bias current, $\gamma_0$. The inclination is proportional to the Lorentz force on the vortex which is induced by the bias current applied to the junction. At last, the total fluxon potential, $\hat{U}_w+\hat{U}_{h_\bot}+\hat{U}_\gamma$, in the presence of both an applied magnetic field and bias current is depicted by the solid line at the top of Fig.~\ref{potentials}. We note that the left well is very shallow and an increment of the bias current would further tilt the potential; a static fluxon pinned in the left well would become unstable and gets trapped in the right well as, in this specific example, it does not have enough energy to move further to the right. The smallest tilting that allows the vortex to escape from a well defines the so-called depinning current, $\gamma_d$. Clearly, the depinning current depends on the applied magnetic field and, in general, for a given field, the depinning currents, $\gamma_{d}^{L}$ and $\gamma_{d}^{R}$, from the left and right wells are different. The deeper is the original potential well from which the fluxon has to escape and the larger is the corresponding depinning current.

\section{The numerical simulations}
 
\begin{figure}[!t]
\centering
\subfigure[ ]{\includegraphics[height=6cm,width=8cm]{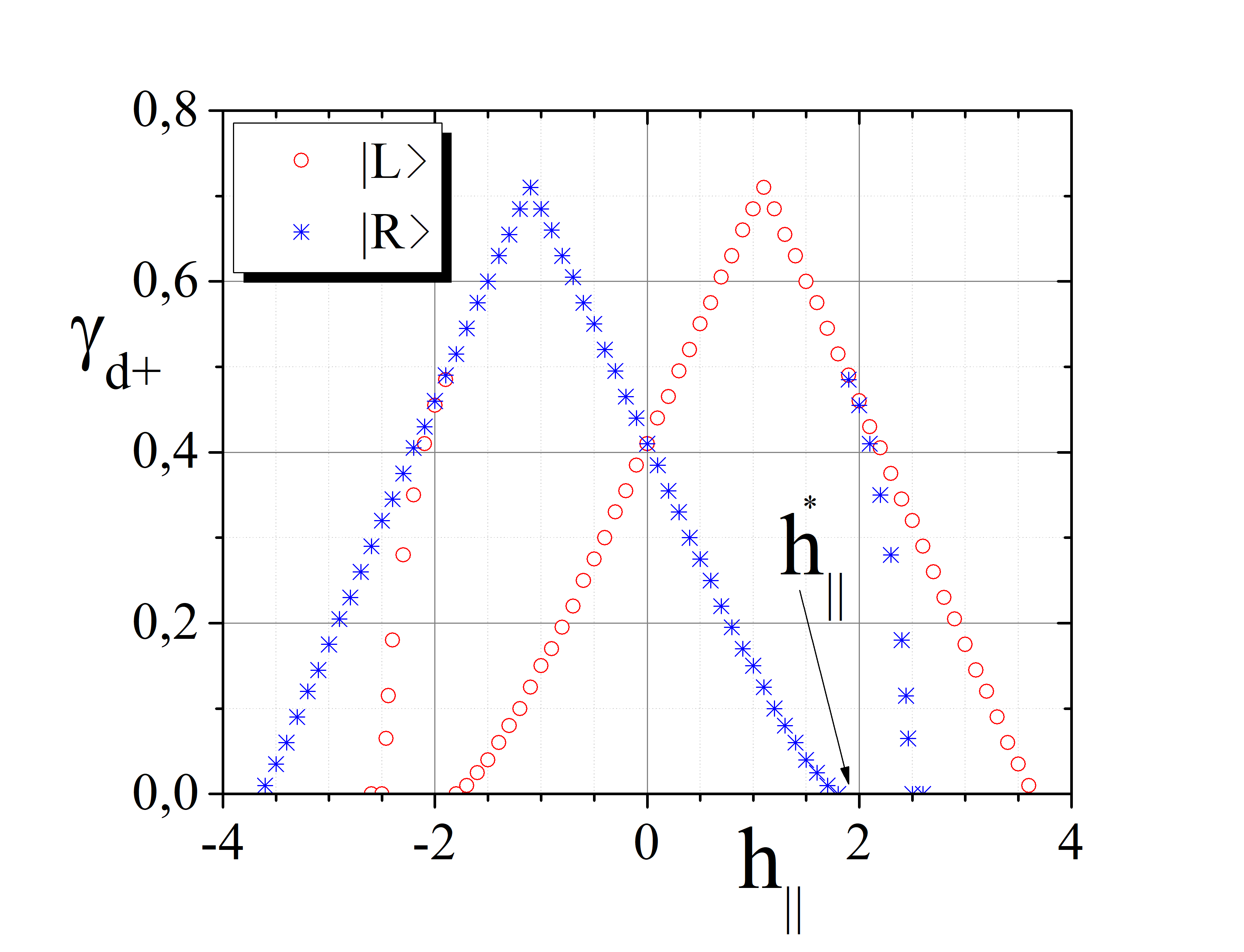}}
\subfigure[ ]{\includegraphics[height=6cm,width=8cm]{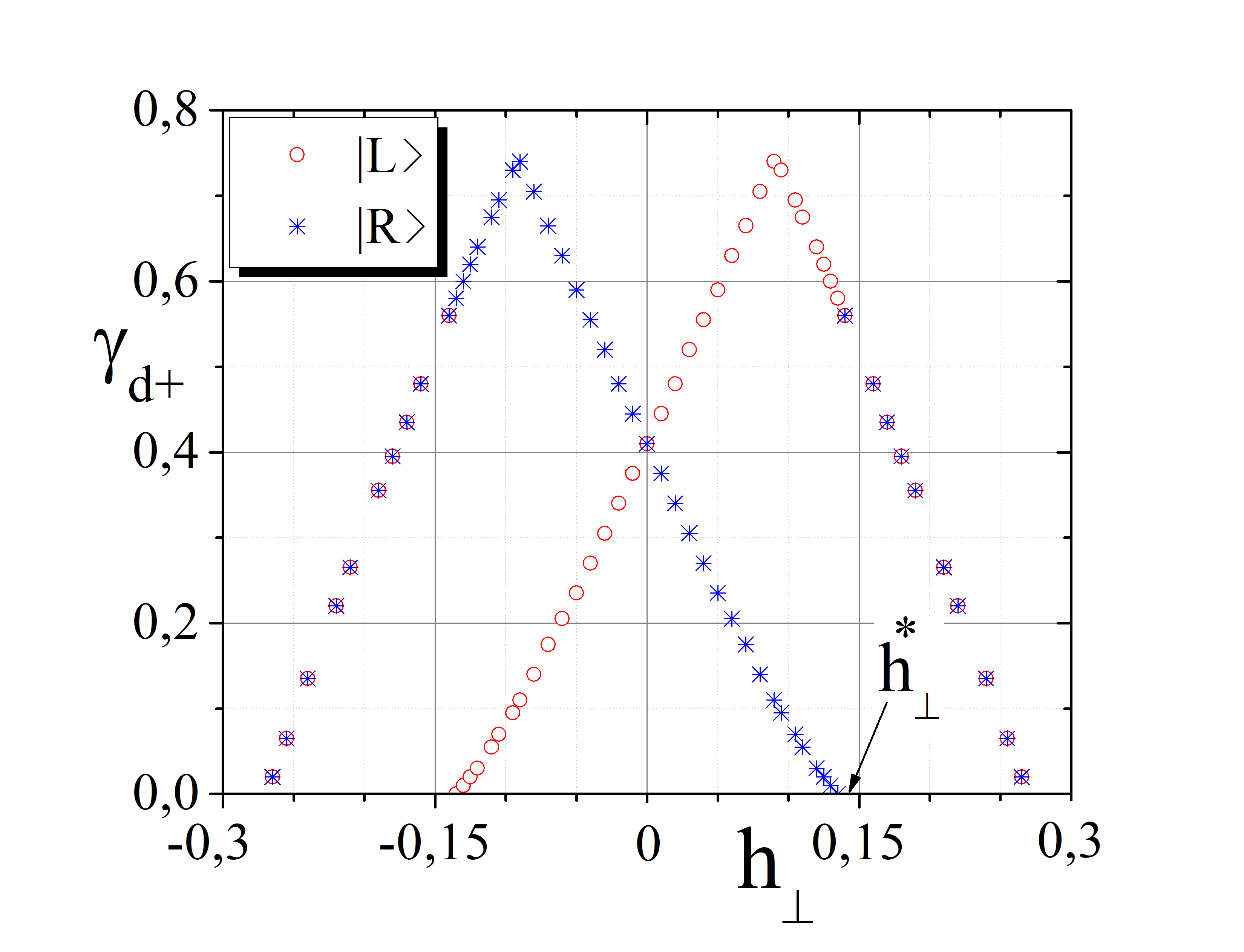}}
\caption{(Color online) Numerically computed field dependencies of the positive fluxon depinning currents of the $|L\rangle$ (open circles for $\gamma_{d+}^{L}$) and $|R\rangle$ (stars for $\gamma_{d+}^{R}$) states for two values of the in-plane field orientation with respect to the annulus major axis: (a) parallel field, $h_{||}$, and (b) perpendicular field, $h_{\bot}$. The magnetic fields are normalized to $J_c c$.}
\label{depinning}
\end{figure}

\noindent In the Figs.~\ref{depinning}(a) and (b) we report the numerically computed field dependencies of the positive left and right depinning currents, $\gamma_{d+}^{L}$ (open circles) and $\gamma_{d+}^{R}$ (crosses), for a (positive) fluxon in the presence of a uniform in-plane magnetic field, respectively, parallel and perpendicular to the CAJTJ's major axis. The numerical simulations of (\ref{psge}) become mandatory whenever the applied bias current and/or magnetic field cannot be considered as perturbations. The commercial finite element simulation package COMSOL MULTIPHYSICS (www.comsol.com) was used to numerically solve (\ref{psge}) subjected to the cyclic boundary conditions in Eqs.(\ref{peri1}) and (\ref{peri2}) with $n=1$.  We set the damping coefficients $\alpha=0.05$ to simulate a weakly underdamped regime and $\beta=0$ as we are considering quasi-static phase solutions. We assumed a uniform current distribution, i.e., $\gamma(\tau)= \gamma_0$. In addition, the field coupling constant, $\Delta$, was set equal to $1$. In order to compare the numerical results with the experimental findings presented in the next section, we set the annulus aspect ratio to $\rho=1/4$ corresponding to a CAJTJ whose largest width is four times larger than the smallest one - as depicted in Fig.~\ref{ConfAnn}(b). Furthermore, the normalized length, $\ell=L/\lambda_J$, was set to be $10\pi$; then, the (smooth) variation of the annulus width occurs over a length, $L/4=2.5 \pi \lambda_J \approx 8 \lambda_J$, quite large compared to the fluxon size. A static fluxon centered either in the left ($\tau_0=-\pi/2$) or right well ($\tau_0=\pi/2$) was chosen for the system initial condition with $\gamma_0=0$ in (\ref{psge}); then the normalized bias current was ramped-up in small adiabatic increments of $0.05$ and the stationary, i.e., time-independent solutions recorded until the fluxon was depinned from its initial state.

\medskip
\noindent We first note that the field dependencies of the depinning current shown in Figs.~\ref{depinning}(a) and (b) are qualitatively similar despite the quite different ways the parallel and perpendicular fields affect the fluxon potential. The most evident discrepancy resides in the magnetic scales and reflects the fact that the junction cross-section seen by a perpendicular field is larger than its parallel counterpart \cite{ekin,nappimonaco}. In both cases the exchange of the fluxon initial position is equivalent to a field reversal, i.e.,  $\gamma_{d+}^{L}(h_{||})=\gamma_{d+}^{R}(-h_{||})$ and $\gamma_{d+}^{L}(h_{\bot})=\gamma_{d+}^{R}(-h_{\bot})$. As expected, the zero-field depinning currents are degenerate, $\gamma_{d+}^{L}(0)=\gamma_{d+}^{R}(0)$, and constiture an appreciable fraction, that is $41\%$, of the zero-field critical current; a smaller value, that is $19\%$, has been reported \cite{SUST18} for a CAJTJ with aspect ratio $\rho=1/2$. These different values reflect the fact that the deeper is the potential well and the largest is the bias current needed to unpin the fluxon. As a parallel magnetic field is turned on, it is seen from Fig.~\ref{depinning}(a) that, the left and right depinning currents change quite linearly and in opposite directions, as long as the amplitude of the applied field is smaller than a characteristic field value $h_{||}^*$ where one of the depinning currents vanishes. In other words, for $|h_{||}|\leq h_{||}^{*}$ the fluxon escape from the $|L\rangle$ and $|R\rangle$ states occurs at quite different bias currents. The state with the higher depinning current corresponds to a deeper potential well when the CAJTJ is unbiased; numerical simulations show that when the fluxon escapes from the state with higher depinning current it starts to travel along the annulus perimeter and switches the junction into a dynamic state with a finite voltage across the junction proportional to the fluxon time-averaged speed. Therefore, in this particular case, the depinning current identifies with the switching current that can be easily determined experimentally. This is not necessarily the case for the state with lower depinning current, as it might happen that, once depinned, the fluxon gets trapped in the opposite well which has an higher depinning current \cite{carapella04}, so that the junction remains in a time-independent, i.e. zero-voltage, state. Numerical analysis shows that this situation only occurs for magnetic fields whose absolute values are close to - but lower that - $h_{||}^{*}$. Furthermore, the re-trapping drastically depends on the junction losses that may dissipate the energy of the depinned fluxon well before it overcomes the opposite well. When the loss parameter $\alpha$ in (\ref{psge}) is decreased to $0.01$ the re-trapping field range shrinks, but does not vanish; therefore, even lower losses should be used in the numerical analysis to investigate the re-trapping conditions. However, great care must be taken to simulate low-damping nonlinear systems, since, besides the longer transients, the results are very sensitive to the numerical algorithm adopted to integrate the partial differential equation. Nevertheless, setting $\alpha=0.05$ the fluxon escaping from the state with the lower depinning range jumps over the opposite well and enters the running mode in a field range of approximately $|h_{||}|\leq h_{||}^{*}/2$. In this range we can talk again of a switching current that coincides with the depinning current. We recall that in the experiments with high-quality JTJs the losses drastically decrease with the temperature \cite{PRB94} and it is not difficult to reach damping parameters as small as $0.001$ \cite{welner85}. Notably, the existence of a  range of parallel fields in which also the lower depinning current turns the CAJTJ into a finite voltage state implies that a switching current measurements \cite{Lisitskiy14} allows to localize the vortex in one of the two states (this technique has been successfully used to prove the existence of a DWP in other Josephson vortex qubit prototypes \cite{wallraff03,carapella04,kemp10}). As $|h_{||}|$ exceeds $h_{||}^{*}$ the unbiased fluxon potential becomes single welled, the information about the vortex initial state is lost and the left and right depinning currents suddenly coincide. At this point an eventual reduction of the field amplitude below $h_{||}^{*}$ and even its full removal leaves the fluxon in the left or in the right well depending on its original sign; in different words, the proper ramping up and down of just the parallel field represents a viable procedure to prepare the vortex state. Similar considerations also apply to the case of a perpendicular magnetic field; the only significant difference is that the preparation of the fluxon state, beside a magnetic field $|h_{\bot}|>h_{\bot}^{*}$, also requires a small bias current that breaks the system symmetry \cite{note}.



\section{The measurements} 

\begin{figure}[!t]
\centering
\subfigure[ ]{\includegraphics[width=7cm]{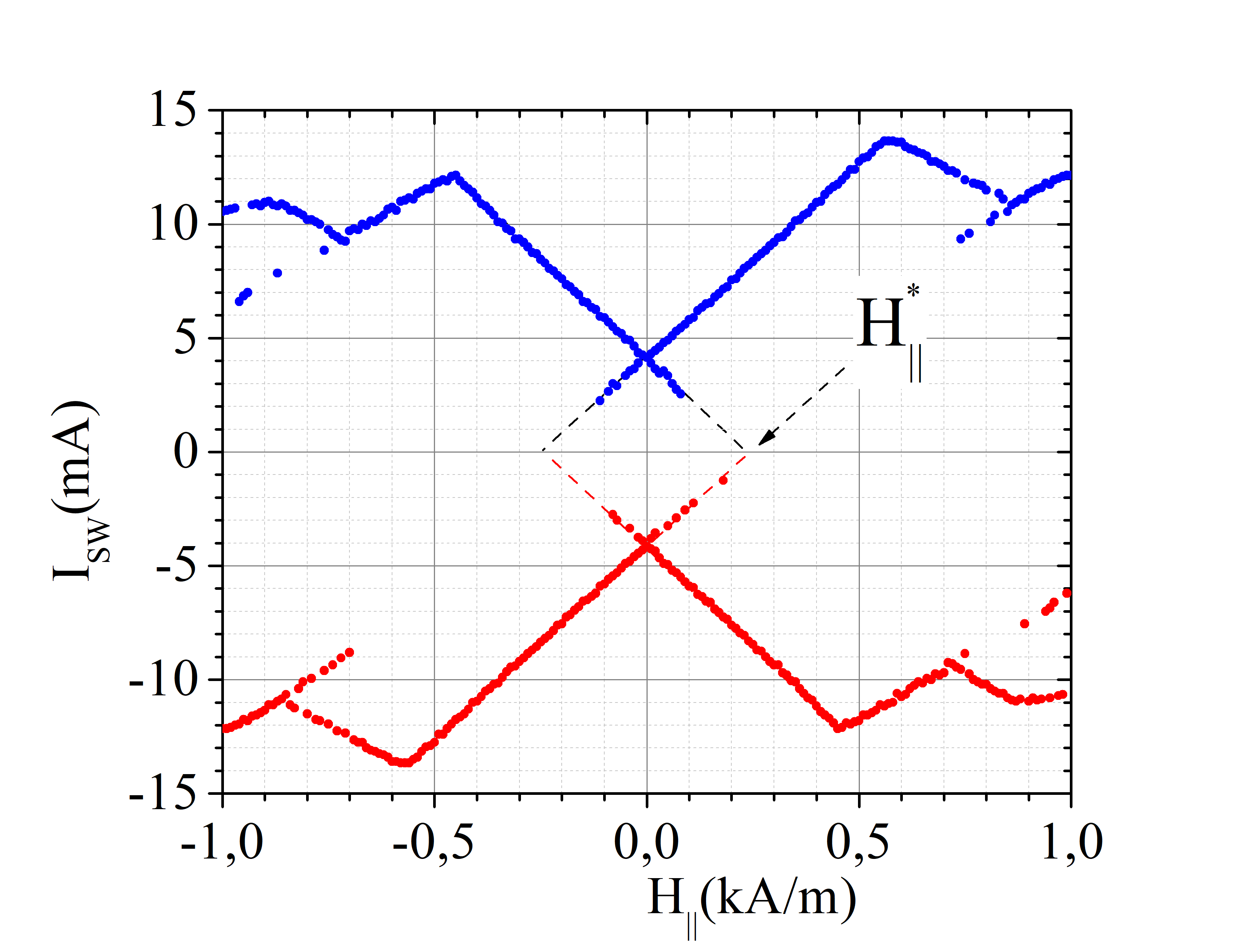}}
\subfigure[ ]{\includegraphics[width=7cm]{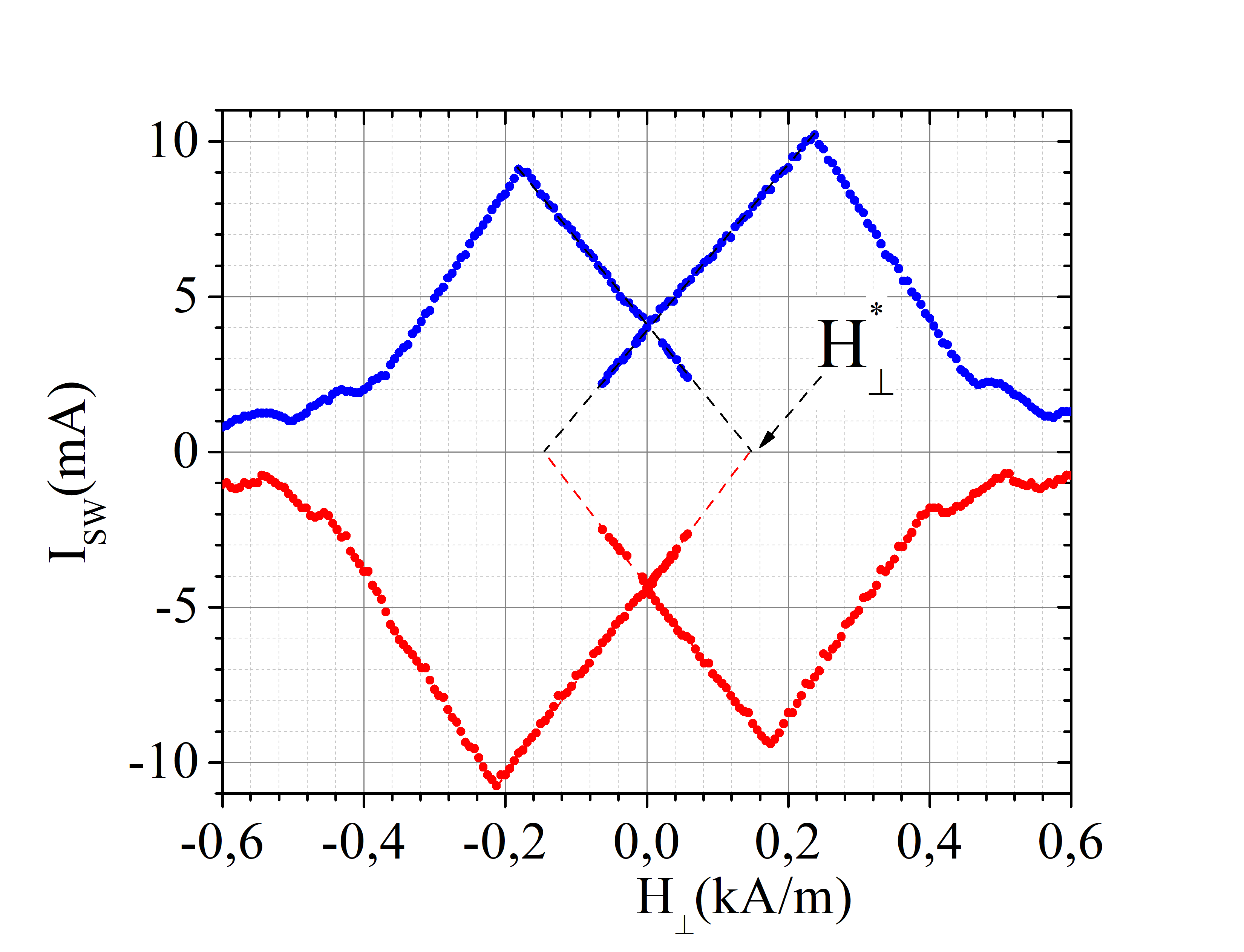}}
\caption{(Color online) Positive and negative switching currents, $I_{SW}$, recorded at $T=4.2\,K$, by continuously sweeping the bias current as an in-plane magnetic field is changed; the dashed lines are the extrapolations of the almost linear branches that help to locate the threshold field $H_{||,\bot}^{*}$; (a) parallel field, $H_{||}$, and (b) perpendicular field, $H_{\bot}$.}
\label{switching}
\end{figure}

In this section we report on the switching currents measured on high-quality window-type $Nb/Al$-$AlOx$-$Al/Nb$ CAJTJs with a single fluxon trapped during the zero-field cooling of the samples through their critical temperature, $T_c\approx 9.2\, K$. This process is known to spontaneously generated one or more fluxons on a statistical basis \cite{PRB06,PRB08} with a probability that increases with the speed of the normal-to-superconducting transition; at the end of each quench the number of trapped fluxons is determined by carefully inspecting the junction current-voltage characteristic (IVC) and measuring the voltage of possible current branches, the so-called \textit{zero-field steps}, associated with the traveling of the Josephson vortices around the annulus. The samples designed to test the theory had the so called \textit{Lyngby-type} geometry\cite{davidson85} that refers to a specularly symmetric configuration, as that shown in Fig.~\ref{SEM}, in which the width of the current carrying electrodes matches one of the axes of the outer ellipse and the tunneling area is obtained by the superposition of two superconducting rings. More details on the samples description end their fabrication were reported, respectively, in Ref.\cite{SUST18} and Ref.\cite{Filippenko}. Some electrical parameters (measured at $4.2\,K$) and the geometrical details of the tunneling area for our CAJTJs are listed in Table I. The samples normalized length is $L/\lambda_J\approx 32.2\approx 10\pi$. We observe that the annulus polar width, $\Delta w_{max}$, slightly exceeds the fluxon size, $\lambda_J$, meaning that the samples are not strictly one-dimensional. In addition, the smallest curvature radius of the master ellipse occurring at the equatorial points, $c \rho \sinh\,\bar{\nu}\approx 2.9 \mu m$, is smaller than the fluxon size, a fact that may induce an interaction (repulsion) between the leading and trailing edges of the fluxon. 

\begin{table}[!t]
\caption{Some electrical parameters (at $4.2\,K$) of our CAJTJs and the geometrical details of their tunneling area.}
		 \label{table}
	\centering
	 
	 	\begin{tabular}{|c|c|c|c|c|c|c|c|c|c|}
		 \hline
     $J_c$ & $\lambda_J$ & $\rho$ & $\bar{\nu}$ &  $\Delta \nu$ & $c$ & $ \Delta w_{min}$ & $\Delta w_{max}$ & $A$ & $L$\\
		\hline
		 $kA/cm^2$ & $\mu m$ & & &  $$ & $\mu m$ & $\mu m$ & $\mu m$ & $\mu m^2$ & $\mu m$\\
		\hline
    	1.1 & 6.2 & 0.25 &0.256&0.18&45.1&2.1&8.4&1310&200\\
		 \hline
				\end{tabular}
\end{table}

	 %

\medskip
\noindent Several nominally identical CAJTJs fabricated on different chips within the same run of a standard $Nb-AlO_x-Nb$ trilayer process were investigated. Their electrodes were either parallel, as shown in Fig.~\ref{SEM}, or perpendicular to the annulus major diameter and they all gave qualitatively similar results regardless of the orientation of the bias. More important was found to be the orientation of the externally applied in-plane magnetic field; in Figs.~\ref{switching}(a) and (b) we report the field dependence of the positive and negative switching currents measured at $T=4.2\,K$ on two representative samples subject to in-plane magnetic fields having orthogonal orientations, respectively, $H_{||}$ and $H_{\bot}$. The switching currents, $I_{SW}$, were obtained by continuously sweeping the bias current with a sufficiently large symmetric triangular waveform with a frequency of few hertz and automatically recording the largest (and smallest) zero-voltage current ten times for each value of the externally applied magnetic field. In agreement with the numerical expectations presented in the previous Section, in a field range centered around zero, the switching current was found to be double-valued. This is a clear indication that the fluxon experiences a potential profile with two stable states and, when the sweeping current crosses zero, the decelerating fluxon stops on a statistical basis in either one of the two different-depth potential wells. The pinning process of a particle slowing down in a DWP drastically depends on the drag force experienced by the particle; in fact, by repeating the measurements at different temperatures, the double-valued field range progressively shrank as the temperature, and so the losses, was increased. Higher temperatures also means larger thermal fluctuations that might induce hopping between the two states. Other noise sources, such as current and/or field  noise, concur on limiting the field range of the fluxon state having the lower switching current; in fact, few more values of the magnetic field were found to show a double-valued switching current using a manual battery-operated ramping of the bias current. The moderate skewness of the switching current plots is ascribed to fact that for both samples the bias current flow occurs in the direction orthogonal to the applied field \cite{SUST13a}; in this configuration the self-field adds to the external field in the second and fourth quadrants, while in the first and third quadrants it partially compensates the applied field. We found the switching current field dependence to be quite linear and the dashed lines in Figs.~\ref{switching}(a) and (b) are the linear extrapolations that help to locate the threshold field $H_{||,\bot}^{*}$, i.e., the largest theoretical absolute value of the field that would yield a double-valued switching current whenever both the dissipation and noises can be neglected. Therefore, according to our measurements, the determination of the fluxon state can be reliably achieved by applying to an unbiased CAJTJ an in-plane magnetic field that is known to have quite different switching currents for the the two states and then incrementing the bias current with a constant rate until a sudden jump of the voltage from zero to any finite value is detected.

\medskip
\noindent It is worthwhile to mention that the finite voltage observed after a switch does not necessarily have to be $v \Phi_0/L$ corresponding to the fluxon continuous motion along the annulus perimeter, $L$, with a certain time-averaged speed $v$. In most cases the junction switched to a state of free-running phase on the McCumber curve, i.e., at a voltage close to the junction gap voltage $V_g\approx 2.8\, mV$, in particular when the switch occurred from the state with the higher switching current. It means that the fluxon potential is too {\it steep} to allow a steady motion; in fact, as the numerical analysis shows, the plasma waves emitted by the leading (trailing) edge of the accelerating (decelerating) fluxon might grow in size as far as they break in a fluxon-antifluxon pair; the process of pair nucleation continues until the system becomes unstable \cite{Pagano88}. Beside the potential gradient, this radiative process heavily depends on several factors, such as the fluxon speed and the annulus perimeter, but again, above all it is determined by the system dissipation. Indeed, for the high aspect-ratio CAJTJs considered in this paper, temperatures larger than $T=4.2\,K$ were needed to observe the zero field steps. This fact indicates that, as the system losses are increased, for a given bias current, the fluxon average speed gets smaller; in turn, the emitted plasma waves not only have a smaller amplitude but also die away faster and then the conditions are restored for a regular stable motion of the fluxon.
 

\section{Conclusion}

A Josephson vortex trapped in a long and narrow CAJTJ in the presence of a in-plane magnetic field is subject to a periodic asymmetric and fine-tunable double well potential accurately modeled by a modified and perturbed one-dimensional sine-Gordon equation. The key ingredient of this potential is the smoothly distributed modulation of the planar tunnel barrier width whose gradient is related one-to-one to the eccentricity of the elliptical annulus. Numerical simulations and experiments carried on high-quality 
$Nb/Al$-$AlOx$-$Al/Nb$ samples with one trapped fluxon demonstrate that a robust two-minima potential can be tailored whose inter-well barrier grows as the annulus is made more eccentric. We have found that the depinning of the fluxon occurs at bias current that, among other things, depends on the potential well  from which the fluxon is escaping. We considered uniform magnetic fields applied in the barrier plane either perpendicular or parallel to the major axis of the CAJTJ. In both cases, although for different reasons, a wide range of magnetic field was observed characterized by two-valued depinning currents, each value corresponding to either one of the potential stable states. Experiments carried out on CAJTJs with an aspect ratio of $4\!:\!1$ indicate that these ranges grow as the temperature, and so the fluxon losses, and/or as the noise sources are reduced.

\noindent The eccentricity of a CAJTJ also has a drastic effect on the motion of a fluxon whose total energy exceeds the potential energy. In fact, as the aspect ratio is increased, the growing accelerations and decelerations the traveling fluxon experiences when approaches the region of, respectively, smallest and larger width are responsible of a periodic radiation of small amplitude waves that destabilize the forward fluxon advancement. Both numerical simulations and experiments with slightly damped CAJTJs having an aspect ratio of $4\!:\!1$ showed no manifestation of the fluxon propagation along the annulus perimeter.   

\noindent Despite all the above considerations, we observed that the eccentricity of the CAJTJs does not affect the robustness and reliability of the operation of a CAJTJ as a Josephson vortex two-state system. Only small quantitative differences were observed in the depinning currents of samples with different aspect ratios. In conclusion, we have demonstrated the full reliability of the procedures for both the preparation and the determination of the vortex in either one of the two potential minima, that are important for the possible realization of a vortex qubit.
 

\section*{Acknowledgment}

\noindent RM and JM acknowledge the support from the Danish Council for Strategic Research under the program the Danish National Research Foundation (bigQ). LVF acknowledges support from the Russian Foundation for Basic Research, grant No. 17-52-12051. 



\begin{thebibliography}{99}
\bibitem{note} A fluxon can be depinned by both a positive or a negative bias current; for the negative depinning currents, $\gamma_{d-}^{L,R}$, it was found that $\gamma_{d-}(h_{||})= -\gamma_{d+}(-h_{||})$ and $\gamma_{d-}(h_{\bot})= -\gamma_{d+}(h_{\bot})$. It follows that the determination of the fluxon state can also be accomplished through the measurement of a negative current switch.
\end{thebibliography}
\end{document}